\documentclass{article}
\usepackage{enumerate,color}
\usepackage{epic,eepic,ecltree}
\usepackage{url}

\catcode`\à=\active \defà{\`a} \catcode`\À=\active \defÀ{\`A}
\catcode`\á=\active \defá{\'a} \catcode`\Á=\active \defÁ{\'A}
\catcode`\ä=\active \defä{\"a} \catcode`\Ä=\active \defÄ{\"A}
\catcode`\â=\active \defâ{\^a} \catcode`\Â=\active \defÂ{\^A}
\catcode`\å=\active \defå{{\aa}} \catcode`\Å=\active \defÅ{{\AA}}
\catcode`\ç=\active \defç{\c{c}} \catcode`\Ç=\active \defÇ{\c{C}}
\catcode`\è=\active \defè{\`e} \catcode`\È=\active \defÈ{\`E}
\catcode`\é=\active \defé{\'e} \catcode`\É=\active \defÉ{\'E}
\catcode`\ë=\active \defë{\"e} \catcode`\Ë=\active \defË{\"E}
\catcode`\ê=\active \defê{\^e} \catcode`\Ê=\active \defÊ{\^E}
\catcode`\ì=\active \defì{\`{\i}} \catcode`\Ì=\active \defÌ{\`{\I}}
\catcode`\í=\active \defí{\'{\i}} \catcode`\Í=\active \defÍ{\'{\I}}
\catcode`\ï=\active \defï{\"{\i}} \catcode`\Ï=\active \defÏ{\"{\I}}
\catcode`\î=\active \defî{\^{\i}} \catcode`\Î=\active \defÎ{\^{\I}}
\catcode`\ò=\active \defò{\`o} \catcode`\Ò=\active \defÒ{\`O}
\catcode`\ó=\active \defó{\'o} \catcode`\Ó=\active \defÓ{\'O}
\catcode`\ö=\active \defö{\"o} \catcode`\Ö=\active \defÖ{\"O}
\catcode`\ô=\active \defô{\^o} \catcode`\Ô=\active \defÔ{\^O}
\catcode`\ù=\active \defù{\`u} \catcode`\Ù=\active \defÙ{\`U}
\catcode`\ú=\active \defú{\'u} \catcode`\Ú=\active \defÚ{\'U}
\catcode`\ü=\active \defü{\"u} \catcode`\Ü=\active \defÜ{\"U}
\catcode`\û=\active \defû{\^u} \catcode`\Û=\active \defÛ{\^U}
\catcode`\ý=\active \defý{\'y} \catcode`\Ý=\active \defÝ{\'Y}
\catcode`\ÿ=\active \defÿ{\"y} \catcode`\˜=\active \def˜{\"Y}
\catcode`\½=\active \def½{!`}
\catcode`\¾=\active \def¾{?`}
\catcode`\ß=\active \defß{{\ss}}
\newcommand{\tmem}[1]{{\em #1\/}}
\definecolor{grey}{rgb}{0.75,0.75,0.75}
\definecolor{orange}{rgb}{1.0,0.5,0.5}
\definecolor{red}{rgb}{1.0,0.0,0.0}
\definecolor{brown}{rgb}{0.5,0.25,0.0}
\definecolor{pink}{rgb}{1.0,0.5,0.5}
\newenvironment{enumeratenumeric}{\begin{enumerate}[1.]}{\end{enumerate}}
\newcommand{\tmstrong}[1]{\textbf{#1}}
\newenvironment{enumerateroman}{\begin{enumerate}[i.]}{\end{enumerate}}
\newenvironment{enumerateromancap}{\begin{enumerate}[I.]}{\end{enumerate}}
\newenvironment{itemizedot}
  {\begin{itemize}}{\end{itemize}}

\title{\huge{Deductive Object Programming} \\
 \large{trying to make object-oriented programming less complex}
}

 \author{François Colonna\\
Laboratoire de Chimie Théorique\\
Université Paris 6\\
email:colonna@lct.jussieu.fr\\
home page:http://galileo.lct.jussieu.fr\\
this document url:\\http://galileo.lct.jussieu.fr/\~{ }frames/ARTICLES/DOP/Provider.tex
}
\begin{document}

\maketitle

\bibliographystyle{alpha}
\begin{center}
{\tmstrong{Preamble}}
\end{center}
This document is a working document, it may never be published in a scientific
journal. It is aimed at starting a discussion on the interest of the kind of
programming method explained below.

Any comments or corrections are welcomed can be written in 
{\color{red}color} in this document and sent back to me.

\begin{abstract}
We propose some slight additions to O-O languages to implement the necessary
features for using Deductive Object Programming (DOP). This way of 
programming based upon the manipulation of the Production Tree of 
the Objects of Interest, result in making Persistent these Objects 
and in sensibly lowering the code complexity.
\end{abstract}

\section{Motivation}

It is a real frustration, when writing some code, not to be able to use the
values of a functionality of an object by simply referencing it. If you do
so, you will obtain an error as soon as you will use the functionality of a
just created object.

In this paper we show that, in fact, this can be achieved quite easily. The
only information necessary to obtain values as soon as an object is
referenced, is its {\tmem{production tree}}. Having remarked that any object
has necessarily been produced by a tree (see below, 
paragraph \ref{ProductionTree}), the trick is therefore to make this tree accessible to the 
{\tmem{programmer}}.

On the other hand, in the usual way of programming, the {\tmem{produced-by}}
relation is never made explicit but hidden in calls to object creation
methods, scattered in the code and difficult to follow. It is therefore a
common experience, when trying to modify somebody else code, to be stuck
during hours or days at the same point because we have no clear idea of
{\tmem{what we have to do before being able to use a given object}}. We
shall show that, by making explicit the {\tmem{produced-by}} relation between
objects, one can avoid to get trapped in this kind of problem : any reference 
to an object can be removed or added to a class without any further remodeling
of the code.

We shall show how deductive programming is implied by this way of managing
objects which may become persistent (see \cite{Persistence}) and can easily be distributed.

We shall give as an example how we have implemented this programming mechanism 
in Eiffel shall conclude by the proposition to add some new features to the 
language to hide the management tools inside the compiler and make the new 
mechanism transparent to the programmer.

\section{The Production Tree}
\label{ProductionTree}

\subsection{definition}

By {\tmem{production tree}} we mean the tree resulting from the relation
{\tmem{produced-by}} between two objects. Contrarily to the {\tmem{client-of}}
relation, the graph generated by the {\tmem{produced-by}} relation cannot
have cycles, otherwise an object could never be produced

This tree defines in a unique way any {\tmem{ground-state}} (see paragraph
\ref{ObjectGroundState}) of a given object.

As the edges of the tree are totally defined by the code, the only degrees of
freedom left to define an object state are the values of the leaves (the {\tmem{parameters}}values). The {\tmem{production tree}} and a set of
coherent {\tmem{parameters}} are nothing else than the 
{\tmem{persistence closure}} of the object, this point will be developed
below (see paragraph \ref{PersistentObjects}). It is therefore sufficient to know
{\tmem{the production tree}} of an object and to give its 
{\tmem{calculation conditions}} {\tmem{(parameters}}) to define an object 
{{\tmem{state before any calculation has been done}}.

Once this simple proposition has been stated, there is no difficulty to make it effective, that is :
\begin{enumeratenumeric}
  \item To make possible the automatic production of any object in any state. 
  
  \item To define {\tmem{a key}} from the {\tmem{production tree}} to access
to the corresponding object {\tmem{state}} (stored in a data-base, for
example) before any calculation has been done. This possibility allows to
{\tmem{reuse the object values not only its functionnalities}}, thus
extending the capabilities of programming with objects.
  
  \item To make unimportant (but easily providable) which path the author has
decided to follow to build some target object {\tmstrong{t}} from
pre-existing objects  {\tmstrong{a}}, {\tmstrong{b, ..., x}}, making the
code modification by an alien programmer easier.
\end{enumeratenumeric}

\subsection{new kinds of attributes}
\label{NewKindsOfAttribute}

In the whole paper we shall speak of {\tmem{attribute}} to designate the
couple (memory-function, memory-attribute).

Let {\tmstrong{an$\_$attribute}} be a memory-function returning some
value of type {\tmstrong{A}} stored at {\tmstrong{an$\_$attribute$\_$memory}}
address, we can write :

\begin{verbatim}
feature {ANY}

an_attribute : A is
  do
    Result := an_attribute_memory
  end -- an_attribute

feature {NONE}

an_attribute_memory : A
\end{verbatim}

The couple ({\tmstrong{an$\_$attribute, an$\_$attribute$\_$memory}}) 
represents what we shall abusively call the attribute {\tmstrong{an$\_$attribute}} of type {\tmstrong{A}}.

Classical OO design concentrates (see OOSC2 \cite{OOSC2}) not on what
attributes a class has
but on {\tmem{what methods a class can offer to manipulate them}}. Of course
we agree with this view, nevertheless it leads to hide the
important question {\tmem{how can I build an instance of this class to use its methods?}}.

As the attributes are supposed not to appear in the interface, their role as
a {\tmem{builder}} or as an {\tmem{internal sub-state}} provider is never
mentionned. Focusing an method for reusability purpose is fine, but
{\tmstrong{methods do not determine the state of an object}}, 
while {\tmstrong{attributes do}}. And, using an object in a given state is what a programmer is
at first concerned with, then he is concerned with applying methods on it.

We do not violate the OO principle to hide attributes. Looking at the example
upper you can see that only {\tmstrong{an$\_$attribute memory-function}} will 
appear in the
interface and not {\tmstrong{an$\_$attribute$\_$memory}} which is hidden, as it
should be.

We will now define two kind of attributes : {\tmem{internal attributes}}
and {\tmem{builder attributes}} or {\tmem{external attributes}}.

\subsubsection{internal attribute}

An {\tmem{internal attribute}} is an attribute calculable from the values of
other attributes of the same object. For example, the {\tmstrong{perimeter}}
of a TRIANGLE (see Annexe B in paragraph \ref{NewImplementationOfClassTriangle}) 
is calculable if we know the {\tmem{position}}s of the 3 {\tmstrong{vertices}}.

\subsubsection{builder attribute or external attribute}

An {\tmem{external or builder attribute}} is an attribute calculated outside
the current object. For example, the 3 {\tmstrong{vertices}} of a TRIANGLE,
(see paragraph \ref{NewImplementationOfClassTriangle}).

These attributes are a source of complexity as they usually need a lot of
information to be built. Our programming mechanism consists in providing these
{\tmem{builder}} attributes in the correct state (attribute of the attribute)
with the only knowledge of the name of the attribute and the name of the state
we want it to be in. For example, in a TRIANGLE  we can directly ask for the
{\tmstrong{vertices}} the state ``position'', using this syntax:

\begin{verbatim}
  needs (vertices(``position''))
  vector_1 := vertices.item (1).position
\end{verbatim}

Which ensures that after the call to the procedure {\tmem{needs}}, the code
can use the value of the position of a vertex as shown. The
{\tmstrong{vector}} \_ 1 variable will be assigned to the correct value.

\subsubsection{parameter attribute : calculation conditions}

We call {\tmem{parameter}} a kind of {\tmem{builder attribute}} not built in
the {\tmem{production tree}}, it has to be provided by the User of the code
(i.e. read from the input). A {\tmem{parameter}} is a leaf of the
{\tmem{production tree}}. Any {\tmem{builder attribute}} of basic type 
(BOOLEAN, INTEGER, REAL, STRING) is necessarily a {\tmem{parameter}} : 
it cannot be built.

We call {\tmem{calculation conditions}} the whole set of {\tmem{parameters}}
for a given {\tmem{production tree}}. They define the 
{\tmem{persistence closure}} (see \cite{OOSC2} page ) 
of the {\tmem{production tree}} and determine completely all the
{\tmem{states}} of any object inside the tree. They are pure basic types 
or collections of basic types.

In the case of a TRIANGLE the {\tmem{parameters}} are the coordinates of the 3
{\tmstrong{vertices}}, i.e. 9 real numbers in a 3-dimensional space.

\subsection{Object state, sub-state and ground-state}
\subsubsection{object state}
\label{ObjectState}

We say that an object is in a {\tmstrong{sub-state}} {\tmstrong{s}} if the
exported attribute {\tmstrong{s}} has been computed. 

\subsubsection{object sub-state}
\label{ObjectSubState}

We shall consider two
types of sub-states corresponding to the two types of attributes already
mentioned in paragraph \ref{NewKindsOfAttribute}, the {\tmem{internals}} and the
{\tmem{externals}}(or {\tmem{builder}} sub-states).

The {\tmstrong{State}} of an object is characterized by the list of its
{\tmstrong{sub-states}}.

The {\tmem{production-tree}} handles only builder sub-state.

\subsubsection{object gound-state}
\label{ObjectGroundState}

We shall say that an object is its {\tmstrong{ground-state}} if
it has all its {\tmem{builders}} built. If an object is in its
{\tmstrong{ground-state}} any of its {\tmem{internals}} state can be built.

\subsubsection{well-built object}
\label{WellBuiltObject}

We propose to say that an object is {\tmem{well built}} if all its 
{\tmem{internals}} sub-states
need the same {\tmem{builders}}. In other words, all {\tmem{internal trees}}
share the same leaves : the  {\tmem{builders}}.

For example, a TRIANGLE with 3 vertices as the only builder is well
built.  If a builder color is added it is not, as a color is not needed to
build the perimeter for instance.

\subsection{cyclic connections}

If an object is produced by a tree, its relations with other objects form a
graph which can even be cyclic. We show here, that this case can also be
managed. Consider the following example :

Object {\tmstrong{a}} has an attribute {\tmstrong{b$\_$in$\_$a}} of type
{\tmstrong{B}}

Object {\tmstrong{b}} has an attribute {\tmstrong{a$\_$in$\_$b}} of type
{\tmstrong{A}}

If - in the code - object {\tmstrong{a}} is created first, then object
{\tmstrong{b}} is a builder of {\tmstrong{A}} , because it is referenced in
{\tmstrong{A}} but not created in {\tmstrong{A}}.

Therefore, {\tmstrong{a}} is created, then {\tmstrong{b}} is created,
{\tmstrong{a$\_$in$\_$b}} is computed, then {\tmstrong{b$\_$in$\_$a}}.

If the situation is the opposite you will have the inverse order of
computations.

So, cyclic connections can also be handled, with the {\tmem{production tree}}
mechanism.

\subsection{The solution we propose : The Object Manager}

How to implement this mechanism in an Object-Oriented language ?
The solution that we have implemented consists in associating an 
{\tmem{Object Manager}} to each object of interest 
(see paragraph \ref{PersistentObjects}). 

\subsubsection{the Object Manager specifications}

An {\tmem{Object Manager}} is an object, biunivocally associated to a 
``real'' object, and  able to answer the programmer's request 
{\tmem{provide me with this object in this sub-state}}.

In some sense it is more than an object and less than an ``agent''.
An {\tmem{Object Manager}} obey to the following requirements :
when asked for providing an {\tmem{external object}} 
object {\tmstrong{an$\_$object}} in state {\tmstrong{s}} it

\begin{enumerateromancap}
  \item tries to retrieve object {\tmstrong{an$\_$object}} in state {\tmstrong{s}} from a data-base
  \begin{enumeratenumeric}
    \item if {\tmstrong{an$\_$object}} in state {\tmstrong{s}} is stored returns {\tmstrong{an$\_$object}}
    
    \item if {\tmstrong{an$\_$object}} in state {\tmstrong{s}} is not stored
    \begin{enumerateroman}
      \item creates the instance {\tmstrong{an$\_$object}}
      
      \item launches the memory-function of {\tmstrong{an$\_$object}} corresponding to the attribute {\tmstrong{s}}.
      
      \item stores {\tmstrong{an$\_$object}} in state {\tmstrong{s}} in data-base.
       
    \end{enumerateroman}
  \end{enumeratenumeric}
  \item returns {\tmstrong{an$\_$object}} in state {\tmstrong{s}}
\end{enumerateromancap}

\subsubsection{How the Object Manager is used now ?}
\label{HowTheObjectManagerIsUsedNow}

We give below an example in Eiffel of the part of class TRIANGLE using 
the {\tmem{Object Manager}} {\tmstrong{triangle$\_$om}} for a TRIANGLE 
(line 1 of code below) :

\begin{verbatim}

1:  triangle_om : TRIANGLE_MANAGER

2:  vertices_memory : ARRAY[POINT]

3:  vertices (sub-state : STRING) : ARRAY[POINT] is
4:  do
6:    if vertices_memory = Void then
7:       vertices_from_key (sub-state)
8:    end    

9:    Result := vertices_memory
10:   ensure
11:     Result = vertices_memory
12: end -- vertices

13: vertices_from_key (sub-state : STRING) is
14:  local
15:    vertices_om : ARRAY_POINT_MANAGER
16:  do
 
17:    vertices_om := triangle_om.child_om_extract (``vertices:ARRAY[POINT]'')
18:    vertices_memory := vertices_om.provided (sub-state)

19:    ensure
      vertices_memory_is_built: vertices_memory /= Void
20: end -- vertices_from_key

21: centroid : POINT
22:  do
23:    if centroid_memory = Void then
24:       centroid_memory_build 
25:    end

26:    Result := centroid_memory
27:  ensure
28:    Result = centroid_memory
29: end -- centroid

30: centroid_memory : POINT

31: centroid_build is
32:  local
33:    vertices_local : like ARRAY[POINT]
34:  do
35:     vertices_local := vertices (``position'')  

36:     create centroid_memory.make
37:     centroid := (vertices.item (1) + vertices.item (2) + vertices.item (3))/3

38:  ensure
39:    centroid_is_built: centroid /= Void    
40: end -- centroid_build

\end{verbatim}

We show upper how the {\tmstrong{centroid}} attributes uses the {\tmstrong{vertices}} attributes as if it were already calculated.
The procedure {\tmem{centroid$\_$build}} (line 31) refers  to
{\tmstrong{vertices}} in sub-state ``position''. This assignment (line 35) launches the execution of the memory-function {\tmstrong{vertices}} (line 3).
The first time the code is executed, {\tmstrong{vertices$\_$memory}} is Void,
the procedure {\tmstrong{vertices$\_$from$\_$key}} is therefore called (line 7 and 13).
It asks the Current's Object Manager (line 17) {\tmstrong{triangle$\_$om}} to 
extract from itself the Object Manager of its son class ARRAY[POINT] (line 15).
This Object Manager {\tmstrong{vertices$\_$om}} provides {\tmstrong{vertices$\_$memory}} in the correct sub-state (line18), i.e. provides of the {\tmstrong{vertices}} positions.

\subsection{Proposed Extensions to the Eiffel language}

Most of the code in paragraph \ref{AnUsualImplementationOfClassTriangle} upper can 
become transparent to the programmer if taken into account by the compiler.
For this, four new keywords have to be added to the Eiffel language. Two new
requirements and two new declaration keywords.

\subsubsection{the {\tmem{needs}} requirement}

To be provided with a needed {\tmem{builder attribute}} in a given sub-state: 

{\tmstrong{needs }}(object$\_$1 (sub-state) ,..., object$\_$n (sub-state)

To be provided with a needed {\tmem{internal attribute}}:

{\tmstrong{needs}} (object$\_$1, ..., object$\_$n)

\subsubsection{the {\tmem{uses}} requirement}

defines the list of building procedure depending on the context

{\tmstrong{uses}} (context$\_$1 : object$\_$1$\_$build, ..., context$\_$n :
object$\_$n$\_$build)

We propose 3 kinds of contexts : build, read and set as shown in table \ref{ta:contexts}

\begin{table}
\begin{center}
\begin{tabular}{l|l}
\hline
  context & procedure suffix\\
\hline
  build   & a $\_$build\\
  read    & a $\_$read\\
  set     & a $\_$set\\
\hline

\end{tabular}
\caption{
\label{ta:contexts}
3 kinds of contexts, and the procedure suffixes associated
}
\end{center}
\end{table}

\subsubsection{the internal keyword}

The syntax looks like :

{\tmstrong{attribute : SOME-TYPE internal (building-procedure)}}

To the type  SOME-TYPE the keyword {\tmem{internal}} is added and the name
of the building-procedure is given between parenthesis.

\subsubsection{the builder keyword}

The syntax looks like :

{\tmstrong{attribute : SOME-TYPE builder (sub-state-name)}}

To the type SOME-TYPE the keyword {\tmem{builder}} is added and the name
of the sub-state to be provided is given between parenthesis.

\subsubsection{How the Object Manager could be implemented ?}
\label{HowTheObjectManagerCouldBeImplemented}

\section{Persistent Objects : a new object sub-category}
\label{PersistentObjects}

Because the {\tmem{ground-state}} (paragraph \ref{ObjectGroundState})
of an object is equivalent to its persistence closure 
(see \cite{OOSC2}, page 252) and
because we have a mechanism allowing {\tmstrong{to define the state of an object before it is created}} it easy to make it persistent and consequently
to check if it has not been stored somewhere (a data-base) in that state.

This is of course not as easy with usual programmation not making the {\tmem{production tree}} explicit. Most of the time, if the object has been
computed during a previous task, you have no mean to know it and the object
has to be recomputed.

Using the {\tmem{Object Manager}} mechanism, it is possible to build a 
{\tmem{key}}
to characterize uniquely any {\tmem{ground-state}} of an object of interest and
to store them in a data-base,  we call {\tmstrong{persistent objects}} 
the objects managed in this way. Sub-keys can also be defined to handle sub-sates.

Here, the {\tmem{builder}} (not {\tmem{creation}}) procedures are
{\tmstrong{closed}} : no further information is needed to invoke them 
(this information is already known by the {\tmem{production tree}}).

In Chapter 8 of his book OOSC2 \cite{OOSC2} at the top of page 236, 
Bertrand Meyer says (although in an other context) :

... {\tmstrong{what you need is, rather than a creation instruction, an  assignment operation that attaches a reference to an already existing object}}.

It is exactly what {\tmem{Persistent Objects}} are aimed to :
create an object in one of its possible states as soon as it is assigned. While what Eiffel propose is to
create an object in an {\tmem{empty}} state or in a unique {\tmem{built}}
state, not made clearly explicit, defined by its creation routine and its class
invariant, this is not enough.

\section{What is lacking in the Class Interface}

An object Class is designed to provide a set of {\tmstrong{functionnalities}}
to manipulate their instances : {\tmem{the interface}}.

Looking at this interface, the question {\tmem{what can I do with it?}} can be
answered, and how to reuse a piece of code already written by somebody else.

However, before using a functionality of a class you have to create an
instance, and {\tmstrong{to create it in such a state as to be sure that this functionality will be usable}} (will have values solving your problem 
not default values). That is precisely
the information lacking in the class interface : 
{\tmem{how to reach the state in which you wish to use one instance, not bothering with all information necessary to reach it?}}.

It is most of the time a complex task to build an instance needed in order to
use it. Why ? Because in usual OO programming the {\tmem{object builder}}
procedures are {\tmem{opened}} that is to say when you invoke them, you
need to feed them {\tmem{from outside}} with some necessary information 
in an argument list,
they do not provide a {\tmstrong{closed object state}}.

In Eiffel you write
\begin{verbatim}
create this_attribute.make (some parameters)
\end{verbatim}

and not

\begin{verbatim}
create this_attribute.make (``some_state'')
\end{verbatim}

For example, if you are writing a new class (for example, 
{\tmstrong{TRIANGLE$\_$PYRAMID}} of paragraph 
\ref{TriangularPyramid})  needing 
an attribute {\tmstrong{base}} of type {\tmstrong{TRIANGLE}} 
to use its {\tmstrong{surface}}, what you need is not
only to know that {\tmstrong{surface}} is of type {\tmstrong{REAL}} and
requires that the {\tmstrong{sides}} (of type {\tmstrong{ARRAY[SEGMENT}}]) 
should be defined. What you need is a mechanism replacing the
{\tmstrong{require}} statement on the {\tmem{necessary existence}} of the 
{\tmstrong{sides}} by the {\tmstrong{effective provision}} of the surface
of{\tmstrong{base}}, i.e. 
a {\tmstrong{REAL}} number which represents its computation, 
{\tmstrong{whenever you make a reference to it}} and whatever the method to built it could be.

By looking at the class interface of {\tmstrong{TRIANGLE}} 
(see Annexe \ref{AnUsualImplementationOfClassTriangle} - the class
of {\tmstrong{a$\_$triangle}} - you have no access to this essential
information, but you {\tmstrong{nee}} it, here is
one of the reason why codes are still complex, even if written according to
the best O-O style. Because authors are {\tmem{not forced}} to make explicit
the route they have decided to follow, leading from one object to its son in
the {\tmem{production tree}}

\subsubsection{Notes}

If you code surface$\_$ build (s1, s2, s3: SEGMENT) you suppose that s1,
s2, s3 are already calculated, 
{\tmstrong{outside surface$\_$build}} which is not.

If you code

\begin{verbatim}
surface_build is
  needs (s1, s2, s3)
  do
  end -- surface_build
\end{verbatim}
you tell the code : if s1, s2, s3 are not yet calculated, calculate them.

The calculation of s1, s2, s3 is done {\tmstrong{inside surface$\_$build}}

\section{Deductive Object Programming}

It was a way of procedural programming which used to be popular in the 
seventies (see references \cite{FinMar-1978},\cite{Les-1978}). It is a top-bottom approach :

Start from the final result you want to reach.

Write the procedure to built it. Then write the procedure building the immediate needed objects. 

Iterate until the data.

This way of programming has been re-actualized for Object Oriented Programming
as follows:
\subsection{definition}

\begin{itemizedot}
  \item start from the final result (the Target of the Task, an Object in some
    state : the centroid of a TRIANGLE)

  \item design objects immediately needed to build this Target (the sons of
    its {\tmem{production tree}}) as {\tmem{function-attributes}} (lines 2 and 3
    of paragraph \ref{HowTheObjectManagerIsUsedNow})
    
  \item in the building procedure of the Target, write a reference reference
    to the son-objects this  make them available in the desired state. (lines 35
    of paragraph \ref{HowTheObjectManagerIsUsedNow})
    
  \item iterate over the objects needed and the objects needed by the objects
    needed until parameters (not buildable but readable objects) are reached,
    read them.
\end{itemizedot}

In other word, instead of building an intermediate object needed to reach your
goal, write your code {\tmstrong{as if these needed objects were already built}}, and defer their building or retrieving from a storage to a specific
object, here the {\tmem{Object Manager}}.

Programming with Persistent Objects or Deductive Object Programming are a same
thing : the programmer never cares of providing values to a 
function-attribute-persistent-object, he declares it
as a builder function-attribute and uses it.

It must be remarked that the sequence of calls to the building routines are
included in each other

a$\_$build calls b$\_$build which calls c$\_$build ... until the leaves

In the usual (bottom-up) way of programming one starts building what is needed 
first (the data-leaves) then climbs up the tree, the sequence of calls in not 
inclusive :

build the leaves when done build x, ... when done build c when done build b
when done build a

and the programmer has to known the whole tree.

\subsection{connection with the Production Tree}

The connection between {\tmstrong{DOP}} and the {\tmem{Production Tree}} is clear
:
both needs to make explicit the sons of the relation {\tmem{built-by}} for any
class to be programmed. The whole {\tmem{Production Tree}} will be built
automatically from this basic information.

As we have already mentioned upper, the {\tmem{production tree}}
is the main information necessary to let the compiler be able to produce an
object in a given sub-state by simple reference (point two of deductive
programming pre-requisite).

\subsection{improvement of code Quality}

\subsubsection{lower complexity}
\label{LowerComplexity}

The complexity is lowered because the only information a programmer has
to known to program a new class is the interface of the son-classes. 
There is no need for him to know the whole {\tmem{production tree}} 
as it is implicit in usual programming.

The complexity is also lowered because the state of the objects are now
made explicit and one can access to the values of any object without taking 
care the complex way to get them.

For example, suppose we want to program a class TRIANGULAR$\_$PYRAMID 
to use the surface of its triangular base, we shall write :

\label{TriangularPyramid}
\begin{verbatim}

class TRIANGULAR_PYRAMID

apex : POINT builder (``position'')

base : TRIANGLE builder (``surface'')

base_surface : REAL is
  needs base(``surface'')
  do
    Result := base.surface
  ensure
    Result = base.surface
end -- base_surface
\end{verbatim}

This is sufficient to obtain the surface of attribute {\tmstrong{base}}
computed according to the context of the Current object 
{\tmstrong{TRIANGULAR$\_$PYRAMID}}. No other knowledge is necessary. 

The compiler will be able to tell the programmer that the
three vertices coordinates of this particular base-triangle have 
to be provided as data
of the Task, as well as the apex coordinates, to defined completely the new 
class instances.

The procedures {\tmstrong{triangle$\_$from$\_$key}} and 
{\tmstrong{apex$\_$from$\_$key}} (similar the that of line 13 of paragraph
\ref{HowTheObjectManagerIsUsedNow} are taken into
account by the compiler.

\subsubsection{objects reusability}

{\tmem{Persistent Objects}} have the property of {\tmstrong{object reusability}}
i.e. :
as objects the code of their Class is reusable, as persistent objects their {\tmstrong{values}} are reusable.

\subsubsection{more evolutionary capabilities}

{\tmem{Persistent objects}} are more independent than usual objects.

If you need to modify a piece of code, that is to say, take out some attribute
and put it elsewhere, the code will re-adapt automatically, because the
{\tmem{production tree is not hard coded}} as in an usual program
but {\tmstrong{dynamically built by the compiler}} from the father-son
couples.

This property is important for the maintainability of always growing systems as
those used in scientific simulations.

\subsubsection{better class design}

A good design implies a few builders and that the
{\tmem{internal attributes}} all share the same 
{\tmem{calculation conditions}}. That is to say all {\tmem{internal trees}} share the same
leaves.
If it is possible to cluster the leaves, because some 
{\tmem{internal attributes}} use only a sub-set of the leaves,  it can be
a sign of a bad design. The class has to be splitted
in one or several heirs, each of them being well-built (see paragraph \ref{WellBuiltObject}.

\subsection{distribution of the building of Objects Persistent}

To manage the distribution of objects building  
first of all one needs to set up 
their {\tmem{production tree}}. This can be a huge task to be done with a 
usual code. 

Using {\tmstrong{DOP}}, it is easy to distribute the building
of the nodes of the {\tmem{production tree}}
on some defined processors of a cluster or the nodes of a grid : this
new functionality can be implemented in the {\tmem{Object Manager}}.

\subsection{extending the Concept of a Calculation}

Instead of computing values we can use the same mechanism to compute whatever
property of an object. For example we can compute the cpu time, the memory,
the disk space to be used, the choice of a given processor or grid node.

The management of the  {\tmem{production tree}} at compilation time, allows
any kind of work flow simulation on the code.

\subsection{iterative processes}

Iterative processes (optimization, Monte Carlo simulation, molecular dynamics)
are very common in scientific calculations.

An iterative process is a process which computes iteratively the same
{\tmem{Target}} object and modifies at each iterations the
{\tmem{calculations conditions}} of this {\tmem{Target}}.

As the {\tmstrong{DOP}} mechanism aims at computing an object in a 
well defined state, modifying
the calculation conditions will modify {\tmem{de facto}} all the objects whose
state is no more consistent with the new calculations conditions. And only
these objects will be recalculated automatically.

We may point out that using {\tmstrong{DOP}} allows a code to optimize any object attribute
against any parameters of the code (this facility is extensively used in the
QMCMOL code \cite{QMCMOL}).

These processes needs to use :

a boolean function like {\tmem{is$\_$not$\_$ready}} instead of the condition
{\tmem{attribute = Void}} (see lines 6 and 23 of paragraph \ref{HowTheObjectManagerIsUsedNow}. This function is true whenever some node
of the production tree of {\tmem{attribute}} has been modified.

an iteration counter

\section{Conclusion}

We have shown that combining deductive programming with making explicit the
production tree of the {\tmem{objects of interest}} (persistent objects)
increases the re-usability and lowers the complexity of codes.

By adding a few functionnalities to any OO language may totally hide the
agent-like mechanism necessary to manage the persistent objects in a today
language like Eiffel. Doing this, the compiler can take an active part in 
the automatic building of any
persistent object in any of its states. This helps considerably the possibility
to modify somebody else's code in full security 
{\tmem{without any deep knowledge of the code}} and allows a simulation of the calculation flow.

Moreover, the objects managed in this way are persistent by construction and
can also have their calculation distributed.

Anyway, this programmation improvement is not yet sufficient to write codes
fully understandable by an alien programmer expert of the domain,
which is the ultimate
goal of programmation. A mechanism 
to make the {\tmem{author's intentions}} immediately understandable, is still lacking.

\section{Acknowledgements} We are grateful 
to Nicole Levy, Parinaz Davari and Francisca Losavio (PRISM, Versailles)
for their contributions
and to Dominique Colnet and Frederic Merizen (Loria, Nancy) 
for their decisive help in the SmartEiffel
implementation. We thank Gilles Blain and Zahia Guessoum 
(Lip6, Paris) for reading the manuscript.

\section{Annexes}

Comparison between the usual and new implementation of class TRIANGLE, what
has changed.

\subsection{A : an usual implementation of class TRIANGLE}
\label{AnUsualImplementationOfClassTriangle}

Below, we give an example of what the interface of class {\tmstrong{TRIANGLE}}
like now :
\begin{verbatim}
class TRIANGLE

feature {ANY}

sides : ARRAY[SEGMENT] 
  require
    vertices_are_defined: vertices /= Void
  ensure
    sides_are_defined: Result /= Void
  end -- sides

centroid : POINT
  require
    vertices_are_defined: vertices /= Void
  ensure
    centroid_is_built: Result /= Void
  end -- centroid

perimeter : REAL
  require
    sides_are_defined: sides /= Void
  ensure
    perimeter_is_built: Result > 0.0
  end -- perimeter

surface : REAL
  require
    sides_are_defined: sides /= Void
  ensure
    surface_is_built: Result > 0.0
  end -- surface

vertices : ARRAY[POINT]

make (points : ARRAY[POINT])
  require
    points_defined: points /= Void
  ensure
    vertices = points
  end -- make

invariant
  vertices_are_built: vertices /= Void

end -- class TRIANGLE

\end{verbatim}

As far as the objects produced by the class ({\tmstrong{surface}},
{\tmstrong{perimeter}}, {\tmstrong{sides}}, {\tmstrong{centroid}}) things are
pretty fine and their relations are clearly described by the assertions:

to compute the {\tmstrong{surface}} you need the {\tmstrong{sides}} and the
{\tmstrong{perimeter}}, to define the {\tmstrong{sides}} you need the
{\tmstrong{vertices}}, to obtain the {\tmstrong{perimeter}} you need the
{\tmstrong{sides}}.

So, supposing you obtain the {\tmstrong{vertices}}, you have no difficulty to
understand how to compute any other property of a {\tmstrong{TRIANGLE}}, just
by looking at the interface.

The problem starts with the {\tmstrong{vertices}}, where do they come from ?
Knowing that they are an array of {\tmstrong{POINT}}, which has to be provided
not Void, does not help to answer the question 
{\tmem{how to obtain the vertices ?}}.

The new implementation below show the answer: the {\tmstrong{vertices}} are a 
{\tmem{builder}}, the {\tmem{Calculation Manager}} will take care of providing 
them in the sub-state ``position'' i.e. with their positions valued as needed for the current calculation.

This TRIANGLE case may seem trivial, when the same mechanism is applied to the 
calculation of a density matrix from a precise kind of wave-function in
quantum  physics, the programming effort stays as low as it is here, which  
is less trivial to do with usual programming.

\subsection{B : a new implementation of class TRIANGLE}
\label{NewImplementationOfClassTriangle}

Below, we give an example of what the interface of class {\tmstrong{TRIANGLE}}
could look like using the Eiffel extensions :

\begin{verbatim}

class TRIANGLE

feature {ANY}

sides : ARRAY[SEGMENT] internal (sides_build)
   needs 
     vertices ("position") 
   ensure
      sides_are_built: Result /= Void
   end -- sides

centroid : POINT internal (centroid_build)
   needs 
     vertices ("position") 
   ensure       
     centroid_is_built: Result /= Void
   end -- centroid

perimeter : REAL internal (perimeter_build)
   needs 
     sides
   ensure
     perimeter_is_built: Result > 0.0
   end -- permeter

surface : REAL internal (surface_build)
   needs 
     perimeter, 
     sides    
   ensure
     surface_is_built: Result > 0.0
   end -- surface

vertices : ARRAY[POINT] builder (``position'')
   ensure     
      vertices_are_defined: Result /= Void 
   end -- vertices

invariant:
  Current /= Void

end -- class TRIANGLE
\end{verbatim}

\begin{itemize}
\item {\tmem{internal}} keyword means that the attribute is internally built by
the procedure in parenthesis.

\item {\tmem{builder}} keyword means the that the attribute has to be externally
built in the sub-state where the attribute {\tmem{position}} of each
{\tmstrong{POINT}} is defined. 
\end{itemize}

We want to emphasize on the fact that the sub-state ``position'' of 
class POINT, is also apparent in this interface of TRIANGLE. Which tells
the programmer in which sub-state a POINT will be used in a TRIANGLE. {\tmstrong{Now, some context appears in the interface}}.

\subsection{C : the external tree of a TRIANGLE}
\label{TheExternalTreeOfTriangle}

\newpage
\begin{bundle}{triangle}
  \chunk{
    \begin{bundle}{vertex 1}
      \chunk{x}
      \chunk{y}
      \chunk{z}
    \end{bundle}
  }
  \chunk{
    \begin{bundle}{vertex 2}
      \chunk{x}
      \chunk{y}
      \chunk{z}
    \end{bundle}
  }
  \chunk{
    \begin{bundle}{vertex 3}
      \chunk{x}
      \chunk{y}
      \chunk{z}
    \end{bundle}
  }
\end{bundle}

\subsection{C : the internal trees of a TRIANGLE}

\subsubsection{C-a : the internal tree of surface}
\label{TheInternalTreeOfSurface}

\begin{bundle}{triangle}
  \chunk{
    \begin{bundle}{surface}
      \chunk{
	\begin{bundle}{sides 1}
	  \chunk{vertex 1}
	  \chunk{vertex 2}
	\end{bundle}
	\begin{bundle}{sides 2}
	  \chunk{vertex 2}
	  \chunk{vertex 3}
	\end{bundle}
	\begin{bundle}{sides 3}
	  \chunk{vertex 3}
	  \chunk{vertex 1}
	\end{bundle}
      }
    \end{bundle}
  }
\end{bundle}

\subsubsection{C-b : the internal tree of centroid}
\label{TheInternalTreeOfCentroid}

\begin{bundle}{triangle}
  \chunk{
    \begin{bundle}{centroid}
	  \chunk{vertex 1}
	  \chunk{vertex 2}
	  \chunk{vertex 3}
    \end{bundle}
  }
\end{bundle}

\newpage
\bibliography{biblio}
\newpage
\tableofcontents

\end{document}